\documentclass[aps,prl,twocolumn,groupedaddress]{revtex4-2}

\usepackage{amssymb}
\usepackage{amsmath}
\usepackage{graphicx}
\usepackage{xcolor}

\begin{document}


\title{Transport of multispecies ion crystals through a junction in an RF Paul trap}


\author{William Cody Burton}
\email{william.burton@quantinuum.com}

\author{Brian Estey}
\author{Ian M. Hoffman}
\author{Abigail R. Perry}
\author{Curtis Volin}
\author{Gabriel Price}
\email{gabriel.price@quantinuum.com}
\affiliation{Quantinuum, 303 S. Technology Ct., Broomfield, Colorado 80021, USA}


\date{\today}

\begin{abstract}
We report on the first demonstration of transport of a multispecies ion crystal through a junction in an RF Paul trap. 
The trap is a two-dimensional surface-electrode trap with an X junction and segmented control electrodes to which 
time-varying voltages are applied to control the shape and position of potential wells above the trap surface.
We transport either a single $^{171}$Yb\textsuperscript{+} ion or a crystal composed of a $^{138}$Ba\textsuperscript{+} ion 
cotrapped with the $^{171}$Yb\textsuperscript{+} ion to any port of the junction. We characterize the motional excitation 
by performing multiple round-trips through the junction and back to the initial well position without cooling. The final excitation is then measured using sideband asymmetry.
For a single $^{171}$Yb\textsuperscript{+} ion, transport with a $4\;\mathrm{m/s}$ average 
speed induces between $0.013\pm0.001$ and $0.014\pm0.001$ quanta of excitation per round trip, depending on the exit port.
For a Ba-Yb crystal, transport at the same speed induces between $0.013\pm0.001$ and 
$0.030\pm0.002$ quanta per round trip of excitation to the axial center of mass mode. 
Excitation in the axial stretch mode ranges from $0.005\pm0.001$ to $0.021\pm0.001$ quanta per round trip.
\end{abstract}


\maketitle


Trapped ions are one of the leading candidate systems for scalable quantum computers~\cite{wineland1998experimental, kaushal2020shuttling}. In a trapped-ion quantum 
computer, quantum information is stored in the internal atomic states of the qubit ions, and multi-qubit gates are usually performed 
by coupling the internal states of two qubits with a common motional mode~\cite{cirac1995quantum}. Qubit connectivity can be achieved through control of trapping potentials created by 
surface-electrode ion traps~\cite{PhysRevLett.96.253003, leibrandt2009demonstration, schulz2008sideband}. 

All-to-all connectivity can be achieved either by including all qubits in a single crystal~\cite{wright2019benchmarking} or by use of a quantum charge coupled device (qCCD) architecture~\cite{kielpinski2002architecture, pino2021demonstration}, where gates are performed on 
small chains of ions, which can be reordered and reconfigured. Current trapped-ion quantum computers~\cite{pino2021demonstration, wright2019benchmarking, pogorelov2021compact} use effectively one-dimensional (linear trap) geometries. In such linear traps, sorting ions between gates 
is done through a combination of potential well splits~\cite{rowe2002transport}, rotations~\cite{splatt2009reordering}, and linear transports~\cite{rowe2002transport, home2009complete}. Sorting on two dimensional grids additionally requires transport through junctions to connect linear sections. There have been demonstrations of ion transport through several junction geometries, 
including T junctions~\cite{hensinger2006t}, wafer-trap~\cite{blakestad2009, decaroli2021Zurich, blakestad2011xjunction} and surface~\cite{wright2013GTRIXjunction, zhang2022optimization} X junctions, and Y junctions~\cite{amini2010scalabletraps, moehring2011SandiaJunction, shu2014GTRIYjunction}. However, there has been no reported effort to transport a multispecies ion crystal through the junction. Doing so allows for simultaneous transport of a sympathetic coolant ion with each qubit, simplifying the sorting algorithm in a large-scale quantum computer.

\begin{figure}
\includegraphics[width=\columnwidth]{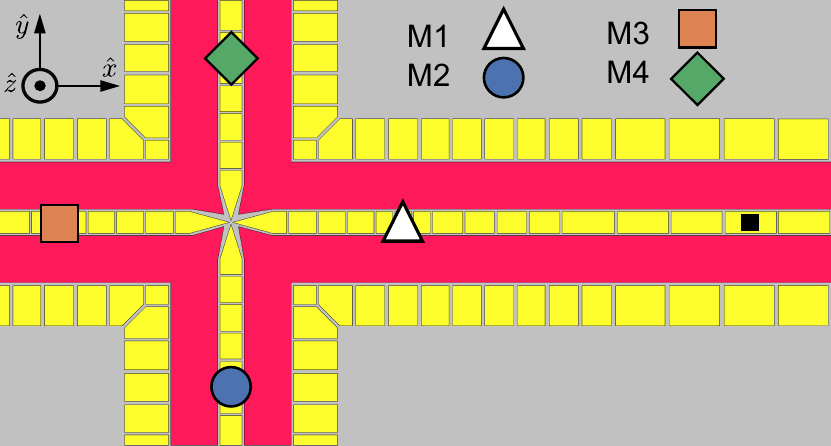}
\caption{\label{trap_fig}Cartoon of the junction trap with the junction center at the origin. The RF rails are in red, and the segmented control electrodes are in yellow.
There are four measurement zones where we can cool and image the ion: M1 (white triangle) at $(375\;\mathrm{\mu m}, 0\;\mathrm{\mu m})$, M2 (blue circle) at $(0\;\mathrm{\mu m}, -375\;\mathrm{\mu m})$, M3 (orange square) at $(-375\;\mathrm{\mu m}, 0\;\mathrm{\mu m})$, and M4 (green diamond) at $(0\;\mathrm{\mu m}, 375\;\mathrm{\mu m})$. For experiments in this paper, we initialized the ions in M1 before performing round-trip transports through the junction to the other measurement zones and back.}
\end{figure}

In this Letter, we present fast, low-excitation junction transport of both single ions and multispecies ion crystals through a microfacbricated surface trap X junction
fabricated at Honeywell (represented in Figure \ref{trap_fig}). 
For single species transport, we used a single $^{171}$Yb\textsuperscript{+} ion, and for multispecies transport, 
we used a crystal composed of one $^{138}$Ba\textsuperscript{+} ion cotrapped with one 
$^{171}$Yb\textsuperscript{+} ion (referred to as a Ba-Yb crystal in the remainder of this paper). For both single species and multispecies transport, we were able 
to transport through the X junction to each of the legs (labeled M1-M4 in Figure \ref{trap_fig}) and back to the starting position with low excitation and without reordering. 
Additionally, we studied the impacts of transport speed, trap electrode drive voltage drifts, and external stray electric fields to ion excitation and survivability.

\begin{table}
\caption{\label{y_excitation_table} Measured axial excitation per round trip of a single ion to the specified end zone and back to M1 at an average speed of $4\;\mathrm{m/s}$. 
The excitation is measured by sideband asymmetry on the axial mode after a variable number of round trips without cooling, and the slope and statistical uncertainty is reported below.}
\begin{ruledtabular}
\begin{tabular}{c c c c}
Crystal & Zone & Round Trip Excitation (quanta) \\
\hline
Yb & M2 & $0.013\pm0.001$\\
Yb & M3 & $0.013\pm0.001$\\
Yb & M4 & $0.014\pm0.001$\\
\end{tabular}
\end{ruledtabular}
\end{table}

Ion surface traps consist of two classes of electrodes: RF rails and segmented control electrodes.
When a radio-frequency (RF) oscillating voltage is applied to the RF rails, charged particles in the vicinity undergo 
micromotion at the RF frequency. This interaction leads to an effective potential $\Phi_{\mathrm{pp}}(\vec{x})$, known as the pseudopotential~\cite{drees1964beschleunigung,wineland1998experimental}.
For each control electrode, we calculate a basis function $\Phi_i(\vec{x})$ 
that describes its contribution to the electrical potential at $\vec{x}$ when 1 V is applied to the electrode. The total potential is given by
\begin{equation}
\Phi (\vec{x}) = \Phi_{\mathrm{pp}}(\vec{x}) + \sum_i V_i \Phi_i(\vec{x}),
\end{equation}
where the $V_i$ are the voltages applied to each control electrode. For a given target trapping well (e.g., a potential 
with a minimum at a defined location and a set of trap frequencies), we can solve for the $V_i$ using 
a constrained optimization method~\cite{blakestad2011xjunction, hucul2008transport}. To create a waveform of voltages that transports an ion between two locations, we solve for a series of
potential wells along the transport path. We then interpolate between these solutions to generate the time-dependent waveform.

The laws of electrostatics place constraints on the sorts of potentials that can be generated. 
We define the total confinement $C = \nabla^2 \Phi \propto \sum_i\omega_i^2$, 
where $\omega_i$ is the frequency of the $i^\text{th}$ normal motional mode of harmonic oscillation of a single ion. Because the control electrodes necessarily produce fields with zero divergence, 
$C \equiv \nabla^2 \Phi_{\mathrm{pp}}$ and depends only on the pseudopotential. In other words, 
no set of control voltages can change the total confinement at a given point -- they can only re-distribute frequency between modes.

\begin{figure}
\includegraphics[width=\columnwidth]{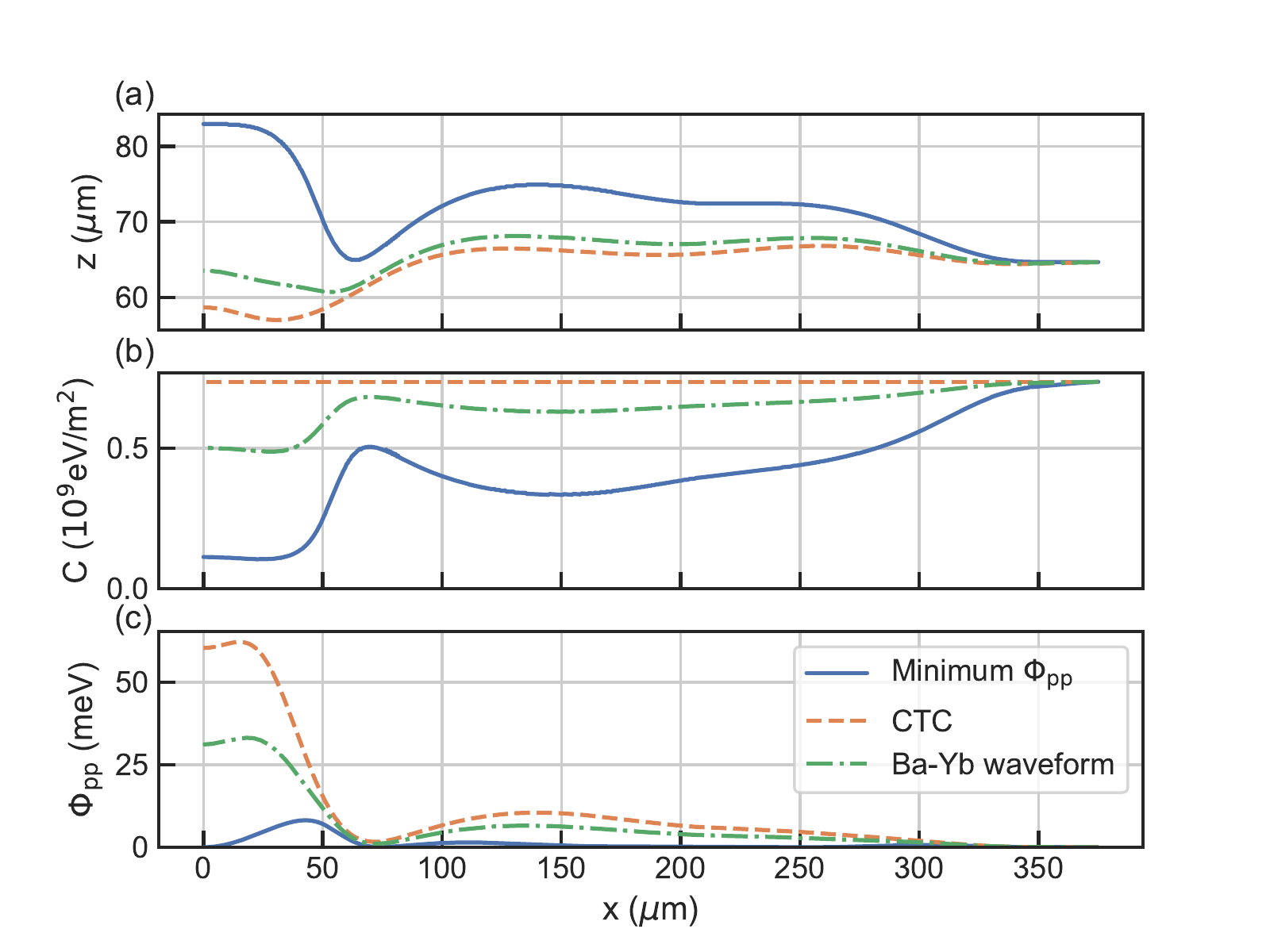}
\caption{\label{ctc_vs_rf_null_fig} Properties of the pseudopotential near the junction center along the path of minimum pseudopotential (solid blue line), 
the path of constant total confinement (CTC) (dashed orange line), and an intermediate path 80\% of the way between the two (dot-dashed green line). 
For a single $^{171}$Yb\textsuperscript{+} ion, we constrain the total potential along the path of CTC, while for a Ba-Yb crystal, we use the intermediate path.
The junction center is at $x = 0$ and the center of zone M1 is at $x = 375\;\mathrm{\mu m}$. (a) The height of each path above the surface of the trap. 
Note that the path of minimum pseudopotential and the path of CTC differ by $\sim20\;\mathrm{\mu m}$ at the junction center. (b) The total confinement ($C$) along 
each path. Along the path of minimum pseudopotential, $C$ drops significantly, while by construction, it does not vary along the path of CTC. (c) Magnitude of the pseudopotential 
($\Phi_\mathrm{pp}$) for a single $^{171}$Yb\textsuperscript{+} ion along each path. The local maximum at around $x = 25\;\mathrm{\mu m}$ is associated
with anti-confinement in the axial direction.
}
\end{figure}

Transport of ions in surface traps have typically followed the path of minimum pseudopotential, which minimizes micromotion and 
sensitivity to noise on the RF source. In an X junction, the total confinement along the pseudopotential minimum near 
the center of the junction drops significantly~\cite{wesenberg2009idealintersections}. 
In this trap, the 
total confinement at the center of the junction is about 1/8 of its maximum value along the path, as seen in Figure \ref{ctc_vs_rf_null_fig}b.

A successful strategy in surface-trap junction transport involves moving the ion off of the path of minimum
pseudopotential~\cite{wright2013GTRIXjunction, shu2014GTRIYjunction}.
We define the path of constant total confinement (CTC) as a path directly below (in $\hat{z}$) the path of minimum pseudopotential, 
with a height that varies to keep $C$ constant along the path. 
We choose the value of $C$ so that the path of CTC and the path of minimum pseudopotential intersect above the center of M1.
Note that this definition of the path of CTC is independent of ion species.
As seen in Figure \ref{ctc_vs_rf_null_fig}a, the path of CTC drops significantly toward the trap relative to the path of minimum pseudopotential.

For single species junction transport, we generated four waveforms that follow the path 
of CTC from the junction center to each of the measurement zones (M1, M2, M3, and M4). We define 
the axial direction to be $\hat{x}$ for M1 and M3, and $\hat{y}$ for M2 and M4 and define the two perpendicular axes to be the radial directions. The 
total confinement along the path permits waveform solutions that maintain a constant axial trap frequency of $1.13\;\mathrm{MHz}$ 
and a large frequency separation of all motional modes over the entire trajectory, preventing transfer of excitation between the different motional modes~\cite{blakestad2011xjunction}. In addition, we created two waveforms that
rotate the principal axes of a single well at the center of the junction by 90 degrees, which adiabatically convert the axial direction from $\hat{x}$ to $\hat{y}$ and back.
The average speed of the transport can be set by linearly scaling the waveform playout in time.

Each single ion transport experiment begins with a $^{171}$Yb\textsuperscript{+} ion trapped in zone M1. We initialize the system 
by performing Doppler and sideband cooling~\cite{PhysRevLett.75.4011} to cool all three motional modes to less than 0.1 quanta of excitation each. 
We then transport the ion to the center of the junction with the time-reversed junction--M1 waveform. To transport to M3, we apply the junction--M3 waveform. 
To transport to M2 or M4, we rotate the principal axes while the well center is stationary at the center of the junction before applying the junction--M2 or junction--M4 waveform. We hold the well center position constant for $1\;\mathrm{\mu s}$ and then
 reverse the transport sequence back to M1. The final excitation is 
measured with Raman sideband asymmetry on the motional mode in the axial direction~\cite{PhysRevLett.75.4011}. To separate non-zero 
initial temperature from excitation due to transport, we measure the excitation as a function of the number of round-trips and extract the 
slope, recorded in Table \ref{y_excitation_table} for transports with average speeds of $4\;\mathrm{m/s}$.

Multispecies crystals present a particular challenge for junction transport because the 
pseudopotential is proportional to the inverse of the ion mass~\cite{drees1964beschleunigung}, while the potential due to control electrodes only depends on the ion charge and is thus common to both 
species. This affects junction transport in two significant ways:
\begin{enumerate}
	\item \label{list_pp_bump} At approximately $25\;\mathrm{\mu m}$ from the center of the junction, there 
is a local maximum of the pseudopotential (see Figure \ref{ctc_vs_rf_null_fig}c), which corresponds to an axial anti-trapping potential. 
When making a transport waveform for a single ion species, our solution method takes this into account and holds the total axial curvature constant.
However for the same waveform, an ion with a different mass will necessarily experience a changing axial curvature, which could lead to excess motional excitation or ion loss.
	\item \label{list_stretching} The minimum of the total potential for an ion of one mass occurs at the location where the gradient of the potential from 
the control electrodes is equal and opposite to the gradient of the pseudopotential. However, the potential minimum for a second ion with a different mass will be at a different position.
\end{enumerate}

Given sufficient degrees of freedom by the control voltages, the total potential for both ion species could be independently controlled. However, this is experimentally impractical 
since the spacing of an ion crystal is generally of the order of a few micrometers while the distance from the trap and the size of the control electrodes are both about $70\;\mathrm{\mu m}$.
Instead, we create multispecies junction transport waveforms through a numerical optimization process. We parametrize possible transport waveforms using two degrees of freedom: the path height parameterized by the fraction of the distance between the CTC  and the pseudopotential minimum, and the potential curvature in the axial direction. These degrees of freedom have several coupled effects:
\begin{enumerate}
	\item A larger axial curvature ensures that the total potential for both species remains trapping even at the peak of the anti-curvature of the pseudopotential, but for a given total confinement it reduces the potential curvature in the radial directions.
	\item The path height affects both the total confinement and the gradient of the pseudopotential at the ion crystal.
	\item The separation of the minima of the total potential for the different ion species is given by a combination of the gradient of the pseudopotential and the curvature in the vertical direction.
\end{enumerate}
Because of the non-trivial coupling, an
exhaustive exploration of the parameter space was employed
to find successful transport waveforms.
At each search point in parameter space, the Ba-Yb waveforms are generated assuming a single synthetic ion with a mass equal to the average ion mass in the Ba-Yb crystal, following the defined transport path, and with a constant total potential curvature in the axial direction. We use a numerical equations-of-motion solver to simulate the behavior of a Ba-Yb crystal during transport in the test waveform and note ion survival and non-adiabatic excitation. We found a broad region of low-excitation waveforms (see Supplemental Material) in parameter space centered around an axial curvature of $9.1\times10^7\;\mathrm{eV/m^2}$ (equivalent to a $1.2\;\mathrm{MHz}$ axial frequency for the synthetic ion) and a path height 80\% of the way between the path of minimum pseudopotential and the path of CTC (dot-dashed green line in Figure \ref{ctc_vs_rf_null_fig}). We use these parameters for multi-species junction transport for the data collected in the remainder of this paper.

\begin{figure}
\includegraphics[width=\columnwidth]{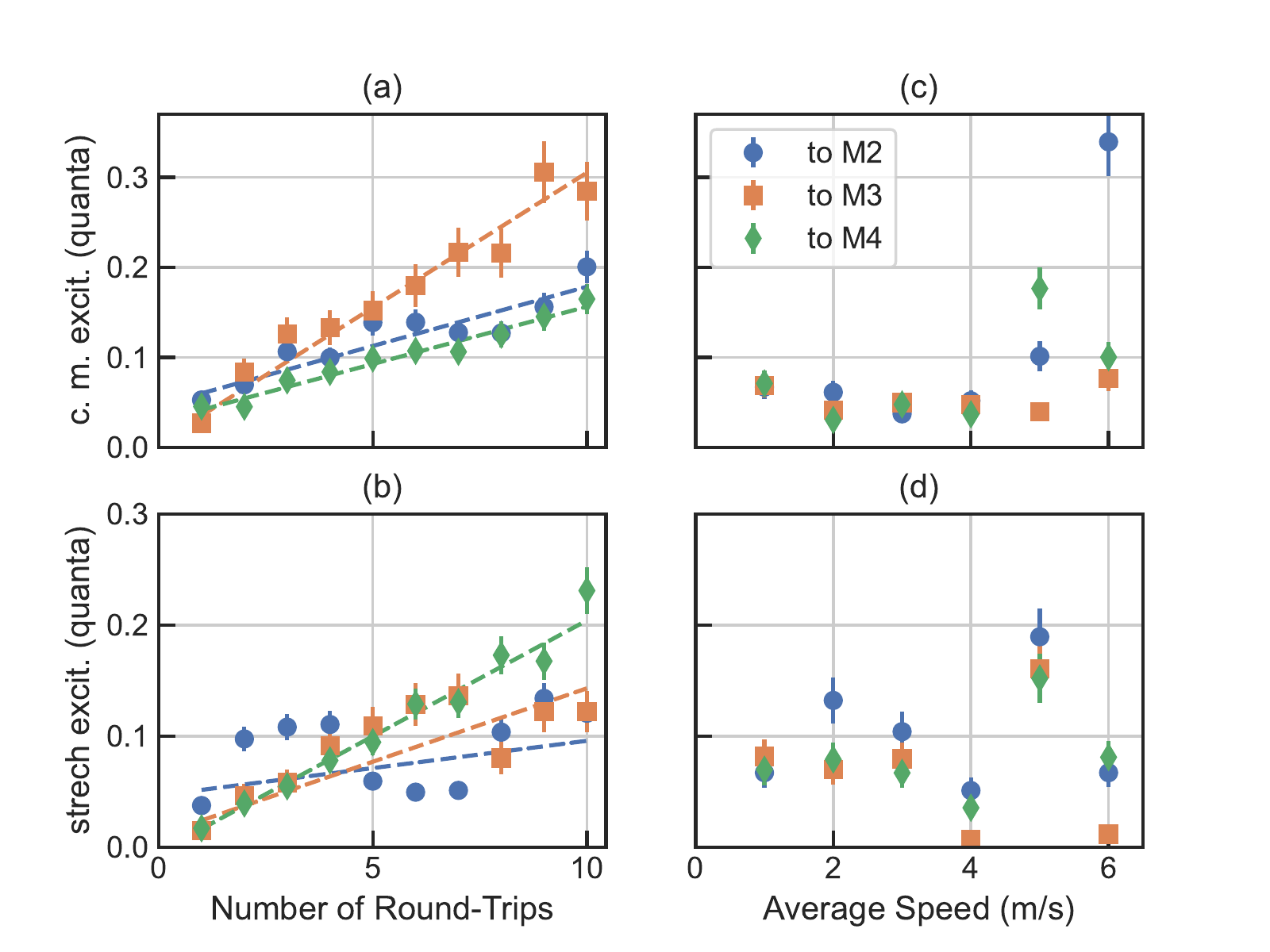}
\caption{\label{yb_data_fig}Measured axial excitation of a Ba-Yb crystal after round trip transports from M1 through the junction to 
M2 (blue circles), M3 (orange squares), M4 (green diamonds), and back. Error bars are $1\sigma$ statistical uncertainty. (a-b) Excitation of the center of mass (c.\,m.) and stretch 
motional modes after a variable number of round trips without cooling. The extracted slopes are reported in Table \ref{yb_excitation_table}. 
(c-d) Excitation of the c.\,m. and stretch modes after one round trip at a variable speed. The background heating rates of $29\pm 4\;\mathrm{quanta/s}$ for the c.m. mode and $3.0\pm 0.5\;\mathrm{quanta/s}$ for the stretch mode are not subtracted from the data.}
\end{figure}

Figure \ref{yb_data_fig} shows the experimental results for multispecies junction transport. We begin each experiment 
with a Ba-Yb crystal trapped in M1, with the $^{171}$Yb\textsuperscript{+} ion closer to the junction. A combination of EIT cooling on the $^{138}$Ba\textsuperscript{+} and 
sideband cooling on the $^{171}$Yb\textsuperscript{+} initilalizes the axial modes with less than 0.05 quanta of excitation, 
and the radial modes with less than 0.3 quanta of excitation. As in the single species transport experiments, we perform a variable 
number of round trips between M1 and the other three zones with an average speed of 4 m/s before measuring the excitation in both 
axial motional modes (center of mass and stretch).
The extracted slopes, reported in Table \ref{yb_excitation_table}, 
indicate that less than or equal to 0.03 quanta of excitation are added per round trip. 
In addition, we  examined the impact of transport speed on induced excitation, with results shown in Figure \ref{yb_data_fig}c-d. We find negligible excitation up to 6 m/s. 
Above this speed, we see evidence of coherent motional excitation in the center of mass (c.\,m.) mode, which is difficult to quantify with sideband asymmetry measurements.
Finally, in the Supplemental Material, we report on the sensitivity of junction transport to drifts in the amplitude of the RF drive and to stray electric fields.

\begin{table}
\caption{\label{yb_excitation_table} Measured axial excitation per round trip of a Ba-Yb crystal to the specified end zone and back to M1 at an average speed of $4\;\mathrm{m/s}$. 
The excitation is measured by sideband asymmetry on the specified axial mode after a variable number of round trips without cooling, and the slope and statistical uncertainty is reported below.}
\begin{ruledtabular}
\begin{tabular}{c c c c}
Crystal & Mode & Zone & Round Trip Excitation (quanta) \\
\hline
Ba-Yb & c.\,m. & M2 & $0.013\pm0.001$\\
Ba-Yb & c.\,m. & M3 & $0.030\pm0.002$\\
Ba-Yb & c.\,m. & M4 & $0.013\pm0.001$\\
Ba-Yb & stretch & M2 & $0.005\pm0.001$\\
Ba-Yb & stretch & M3 & $0.013\pm0.001$\\
Ba-Yb & stretch & M4 & $0.021\pm0.001$\\
\end{tabular}
\end{ruledtabular}
\end{table}

\begin{figure}
\includegraphics[width=\columnwidth]{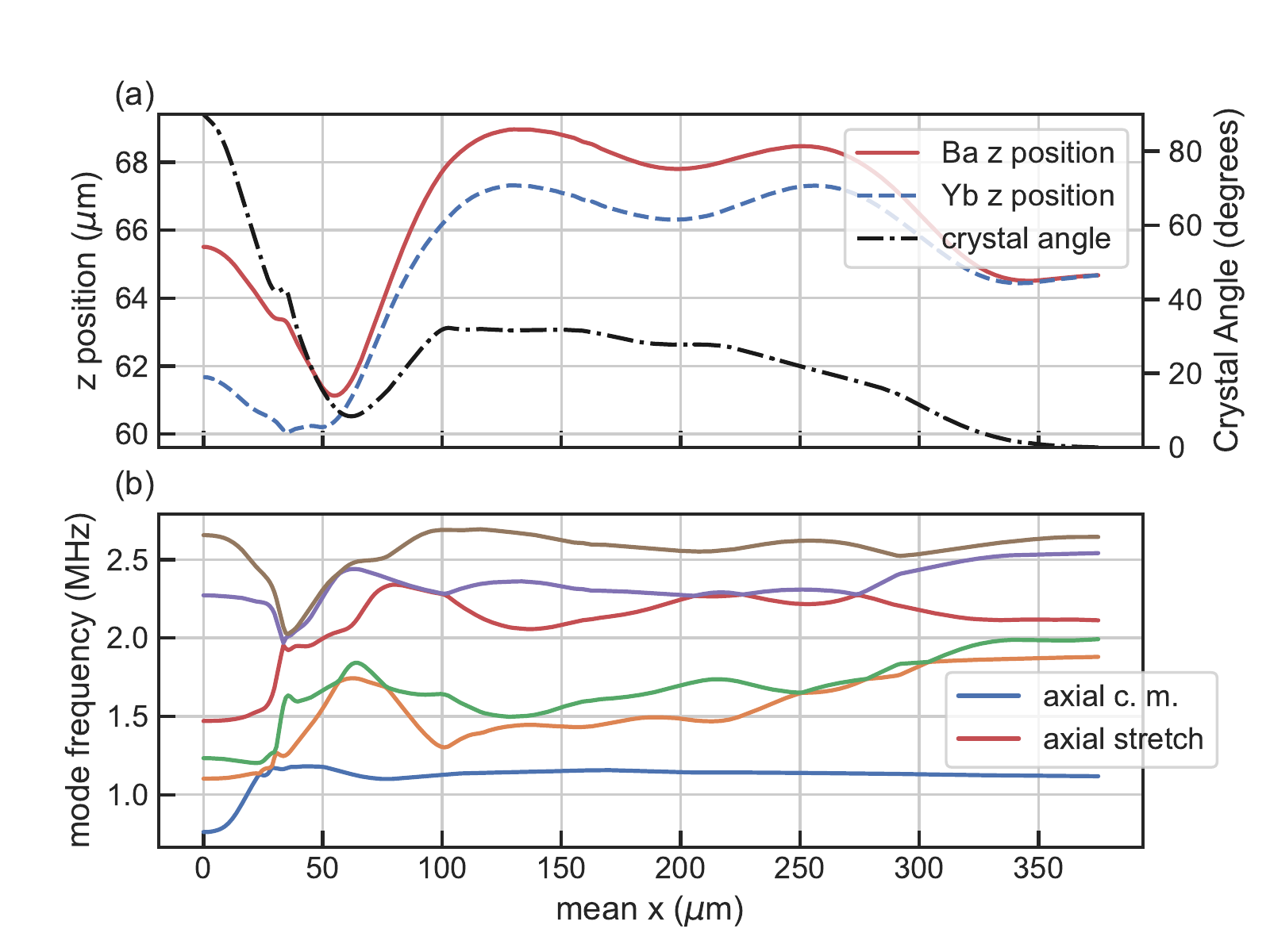}
\caption{\label{yb_freqs_fig} \label{yb_rotation_fig} Predicted properties of the Ba-Yb crystal during junction transport from M1 ($x = 375\;\mathrm{\mu m}$) 
to the junction center ($x = 0$) determined by numerical modeling of the equations of motion. (a) The left axis is the equlibrium height of the $^{138}$Ba\textsuperscript{+} (solid red line) and $^{171}$Yb\textsuperscript{+} 
ion (dashed blue line) above the trap surface, plotted versus the average $x$ position of the two ions. The right axis is the angle of the crystal rotation (black dot-dashed line). The crystal starts in M1 with the $^{171}$Yb\textsuperscript{+} facing the junction and the $^{138}$Ba\textsuperscript{+}
facing away (0 degrees in the $xz$ plane). At the junction center, the crystal is oriented perpendicular to the trap surface (90 degrees). (b) Frequencies of the six normal motional 
modes of the crystal along the transport path. The axial modes in M1 are labeled. Several mode crossings during transport are predicted. 
}
\end{figure}

We numerically simulate the properties of a Ba-Yb crystal using the pseudopotential separately scaled by mass for each ion species.
In Figure \ref{yb_freqs_fig}a, we plot the simulated equilibrium positions of a Ba-Yb crystal during transport from M1 
to the junction center. (The qualitative features of this waveform are shared by all successful Ba-Yb waveforms found in the numerical optimization.) When the crystal is near M1, 
where the trapping potential is similar to that generated in a linear surface trap, the crystal is oriented in the axial direction with the $^{171}$Yb\textsuperscript{+} ion closer to the junction. As the crystal approaches the center of the junction,
due to different contributions to the total potential from the pseudopotential, it rotates to be perpendicular to the trap (the $\hat{z}$ direction), regardless of the starting leg. When the crystal moves away from the junction center, either reversing its motion or moving into one of the other three legs, the crystal rotates back to a horizontal orientation, but always with the $^{171}$Yb\textsuperscript{+} 
closer to the junction, without an explicit rotation waveform. In Figure \ref{yb_freqs_fig}b, we plot the normal mode frequencies versus ion crystal location. Near $x = 25\;\mathrm{\mu m}$,
there are several normal mode crossings that couple some of the radial modes to the axial modes during the transport, complicating analysis of transport performed without pre-cooling the radial modes. In experiments where we pre-cool all motional modes, we see no evidence of excitation leaking into the measured axial modes and can conclude that the motional excitation in the radial modes is low.

In this Letter, we have presented the first implementation of junction transport of a multispecies ion crystal, 
which provides an important tool for minimizing circuit times in a two-dimensional ion trap quantum computer. 
We find that the transport adds very little excitation to all axial motional degrees of freedom at average speeds up to $6\;\mathrm{m/s}$. 
In future work, further analysis of the waveform generation method and numerical optimization of the waveform may allow us to control or eliminate the 
crossing of motional modes during transport or to increase the speed of transport without additional coherent motional excitation~\cite{palmero2014fasttransport}. 
In addition, to demonstrate scalability, we plan to develop the parallel transport of multiple ion crystals through neighboring junctions.

\begin{acknowledgments}
The authors would like to acknowledge Jonathan Andreasen at the Georgia Tech Research Institue for assistance with the surface trap design and Steven Moses at Quantinuum for helpful comments on the text. We thank the entire Quantinuum team for many additional techical and supporting contributions.
\end{acknowledgments}

\bibliography{references}

\begin{thebibliography}{27}%
\makeatletter
\providecommand \@ifxundefined [1]{%
 \@ifx{#1\undefined}
}%
\providecommand \@ifnum [1]{%
 \ifnum #1\expandafter \@firstoftwo
 \else \expandafter \@secondoftwo
 \fi
}%
\providecommand \@ifx [1]{%
 \ifx #1\expandafter \@firstoftwo
 \else \expandafter \@secondoftwo
 \fi
}%
\providecommand \natexlab [1]{#1}%
\providecommand \enquote  [1]{``#1''}%
\providecommand \bibnamefont  [1]{#1}%
\providecommand \bibfnamefont [1]{#1}%
\providecommand \citenamefont [1]{#1}%
\providecommand \href@noop [0]{\@secondoftwo}%
\providecommand \href [0]{\begingroup \@sanitize@url \@href}%
\providecommand \@href[1]{\@@startlink{#1}\@@href}%
\providecommand \@@href[1]{\endgroup#1\@@endlink}%
\providecommand \@sanitize@url [0]{\catcode `\\12\catcode `\$12\catcode
  `\&12\catcode `\#12\catcode `\^12\catcode `\_12\catcode `\%12\relax}%
\providecommand \@@startlink[1]{}%
\providecommand \@@endlink[0]{}%
\providecommand \url  [0]{\begingroup\@sanitize@url \@url }%
\providecommand \@url [1]{\endgroup\@href {#1}{\urlprefix }}%
\providecommand \urlprefix  [0]{URL }%
\providecommand \Eprint [0]{\href }%
\providecommand \doibase [0]{https://doi.org/}%
\providecommand \selectlanguage [0]{\@gobble}%
\providecommand \bibinfo  [0]{\@secondoftwo}%
\providecommand \bibfield  [0]{\@secondoftwo}%
\providecommand \translation [1]{[#1]}%
\providecommand \BibitemOpen [0]{}%
\providecommand \bibitemStop [0]{}%
\providecommand \bibitemNoStop [0]{.\EOS\space}%
\providecommand \EOS [0]{\spacefactor3000\relax}%
\providecommand \BibitemShut  [1]{\csname bibitem#1\endcsname}%
\let\auto@bib@innerbib\@empty
\bibitem [{\citenamefont {Wineland}\ \emph {et~al.}(1998)\citenamefont
  {Wineland}, \citenamefont {Monroe}, \citenamefont {Itano}, \citenamefont
  {Leibfried}, \citenamefont {King},\ and\ \citenamefont
  {Meekhof}}]{wineland1998experimental}%
  \BibitemOpen
  \bibfield  {author} {\bibinfo {author} {\bibfnamefont {D.~J.}\ \bibnamefont
  {Wineland}}, \bibinfo {author} {\bibfnamefont {C.}~\bibnamefont {Monroe}},
  \bibinfo {author} {\bibfnamefont {W.~M.}\ \bibnamefont {Itano}}, \bibinfo
  {author} {\bibfnamefont {D.}~\bibnamefont {Leibfried}}, \bibinfo {author}
  {\bibfnamefont {B.~E.}\ \bibnamefont {King}},\ and\ \bibinfo {author}
  {\bibfnamefont {D.~M.}\ \bibnamefont {Meekhof}},\ }\bibfield  {title}
  {\bibinfo {title} {Experimental issues in coherent quantum-state manipulation
  of trapped atomic ions},\ }\href@noop {} {\bibfield  {journal} {\bibinfo
  {journal} {Journal of research of the National Institute of Standards and
  Technology}\ }\textbf {\bibinfo {volume} {103}},\ \bibinfo {pages} {259}
  (\bibinfo {year} {1998})}\BibitemShut {NoStop}%
\bibitem [{\citenamefont {Kaushal}\ \emph {et~al.}(2020)\citenamefont
  {Kaushal}, \citenamefont {Lekitsch}, \citenamefont {Stahl}, \citenamefont
  {Hilder}, \citenamefont {Pijn}, \citenamefont {Schmiegelow}, \citenamefont
  {Bermudez}, \citenamefont {M{\"u}ller}, \citenamefont {Schmidt-Kaler},\ and\
  \citenamefont {Poschinger}}]{kaushal2020shuttling}%
  \BibitemOpen
  \bibfield  {author} {\bibinfo {author} {\bibfnamefont {V.}~\bibnamefont
  {Kaushal}}, \bibinfo {author} {\bibfnamefont {B.}~\bibnamefont {Lekitsch}},
  \bibinfo {author} {\bibfnamefont {A.}~\bibnamefont {Stahl}}, \bibinfo
  {author} {\bibfnamefont {J.}~\bibnamefont {Hilder}}, \bibinfo {author}
  {\bibfnamefont {D.}~\bibnamefont {Pijn}}, \bibinfo {author} {\bibfnamefont
  {C.}~\bibnamefont {Schmiegelow}}, \bibinfo {author} {\bibfnamefont
  {A.}~\bibnamefont {Bermudez}}, \bibinfo {author} {\bibfnamefont
  {M.}~\bibnamefont {M{\"u}ller}}, \bibinfo {author} {\bibfnamefont
  {F.}~\bibnamefont {Schmidt-Kaler}},\ and\ \bibinfo {author} {\bibfnamefont
  {U.}~\bibnamefont {Poschinger}},\ }\bibfield  {title} {\bibinfo {title}
  {Shuttling-based trapped-ion quantum information processing},\ }\href@noop {}
  {\bibfield  {journal} {\bibinfo  {journal} {AVS Quantum Science}\ }\textbf
  {\bibinfo {volume} {2}},\ \bibinfo {pages} {014101} (\bibinfo {year}
  {2020})}\BibitemShut {NoStop}%
\bibitem [{\citenamefont {Cirac}\ and\ \citenamefont
  {Zoller}(1995)}]{cirac1995quantum}%
  \BibitemOpen
  \bibfield  {author} {\bibinfo {author} {\bibfnamefont {J.~I.}\ \bibnamefont
  {Cirac}}\ and\ \bibinfo {author} {\bibfnamefont {P.}~\bibnamefont {Zoller}},\
  }\bibfield  {title} {\bibinfo {title} {Quantum computations with cold trapped
  ions},\ }\href {https://doi.org/10.1103/PhysRevLett.74.4091} {\bibfield
  {journal} {\bibinfo  {journal} {Phys. Rev. Lett.}\ }\textbf {\bibinfo
  {volume} {74}},\ \bibinfo {pages} {4091} (\bibinfo {year}
  {1995})}\BibitemShut {NoStop}%
\bibitem [{\citenamefont {Seidelin}\ \emph {et~al.}(2006)\citenamefont
  {Seidelin}, \citenamefont {Chiaverini}, \citenamefont {Reichle},
  \citenamefont {Bollinger}, \citenamefont {Leibfried}, \citenamefont
  {Britton}, \citenamefont {Wesenberg}, \citenamefont {Blakestad},
  \citenamefont {Epstein}, \citenamefont {Hume}, \citenamefont {Itano},
  \citenamefont {Jost}, \citenamefont {Langer}, \citenamefont {Ozeri},
  \citenamefont {Shiga},\ and\ \citenamefont
  {Wineland}}]{PhysRevLett.96.253003}%
  \BibitemOpen
  \bibfield  {author} {\bibinfo {author} {\bibfnamefont {S.}~\bibnamefont
  {Seidelin}}, \bibinfo {author} {\bibfnamefont {J.}~\bibnamefont
  {Chiaverini}}, \bibinfo {author} {\bibfnamefont {R.}~\bibnamefont {Reichle}},
  \bibinfo {author} {\bibfnamefont {J.~J.}\ \bibnamefont {Bollinger}}, \bibinfo
  {author} {\bibfnamefont {D.}~\bibnamefont {Leibfried}}, \bibinfo {author}
  {\bibfnamefont {J.}~\bibnamefont {Britton}}, \bibinfo {author} {\bibfnamefont
  {J.~H.}\ \bibnamefont {Wesenberg}}, \bibinfo {author} {\bibfnamefont {R.~B.}\
  \bibnamefont {Blakestad}}, \bibinfo {author} {\bibfnamefont {R.~J.}\
  \bibnamefont {Epstein}}, \bibinfo {author} {\bibfnamefont {D.~B.}\
  \bibnamefont {Hume}}, \bibinfo {author} {\bibfnamefont {W.~M.}\ \bibnamefont
  {Itano}}, \bibinfo {author} {\bibfnamefont {J.~D.}\ \bibnamefont {Jost}},
  \bibinfo {author} {\bibfnamefont {C.}~\bibnamefont {Langer}}, \bibinfo
  {author} {\bibfnamefont {R.}~\bibnamefont {Ozeri}}, \bibinfo {author}
  {\bibfnamefont {N.}~\bibnamefont {Shiga}},\ and\ \bibinfo {author}
  {\bibfnamefont {D.~J.}\ \bibnamefont {Wineland}},\ }\bibfield  {title}
  {\bibinfo {title} {Microfabricated surface-electrode ion trap for scalable
  quantum information processing},\ }\href
  {https://doi.org/10.1103/PhysRevLett.96.253003} {\bibfield  {journal}
  {\bibinfo  {journal} {Phys. Rev. Lett.}\ }\textbf {\bibinfo {volume} {96}},\
  \bibinfo {pages} {253003} (\bibinfo {year} {2006})}\BibitemShut {NoStop}%
\bibitem [{\citenamefont {Leibrandt}\ \emph {et~al.}(2009)\citenamefont
  {Leibrandt}, \citenamefont {Labaziewicz}, \citenamefont {Clark},
  \citenamefont {Chuang}, \citenamefont {Epstein}, \citenamefont {Ospelkaus},
  \citenamefont {Wesenberg}, \citenamefont {Bollinger}, \citenamefont
  {Leibfried}, \citenamefont {Wineland} \emph
  {et~al.}}]{leibrandt2009demonstration}%
  \BibitemOpen
  \bibfield  {author} {\bibinfo {author} {\bibfnamefont {D.}~\bibnamefont
  {Leibrandt}}, \bibinfo {author} {\bibfnamefont {J.}~\bibnamefont
  {Labaziewicz}}, \bibinfo {author} {\bibfnamefont {R.}~\bibnamefont {Clark}},
  \bibinfo {author} {\bibfnamefont {I.}~\bibnamefont {Chuang}}, \bibinfo
  {author} {\bibfnamefont {R.}~\bibnamefont {Epstein}}, \bibinfo {author}
  {\bibfnamefont {C.}~\bibnamefont {Ospelkaus}}, \bibinfo {author}
  {\bibfnamefont {J.}~\bibnamefont {Wesenberg}}, \bibinfo {author}
  {\bibfnamefont {J.}~\bibnamefont {Bollinger}}, \bibinfo {author}
  {\bibfnamefont {D.}~\bibnamefont {Leibfried}}, \bibinfo {author}
  {\bibfnamefont {D.}~\bibnamefont {Wineland}}, \emph {et~al.},\ }\bibfield
  {title} {\bibinfo {title} {Demonstration of a scalable, multiplexed ion trap
  for quantum information processing},\ }\href@noop {} {\bibfield  {journal}
  {\bibinfo  {journal} {Quantum Information \& Computation}\ }\textbf {\bibinfo
  {volume} {9}},\ \bibinfo {pages} {901} (\bibinfo {year} {2009})}\BibitemShut
  {NoStop}%
\bibitem [{\citenamefont {Schulz}\ \emph {et~al.}(2008)\citenamefont {Schulz},
  \citenamefont {Poschinger}, \citenamefont {Ziesel},\ and\ \citenamefont
  {Schmidt-Kaler}}]{schulz2008sideband}%
  \BibitemOpen
  \bibfield  {author} {\bibinfo {author} {\bibfnamefont {S.~A.}\ \bibnamefont
  {Schulz}}, \bibinfo {author} {\bibfnamefont {U.}~\bibnamefont {Poschinger}},
  \bibinfo {author} {\bibfnamefont {F.}~\bibnamefont {Ziesel}},\ and\ \bibinfo
  {author} {\bibfnamefont {F.}~\bibnamefont {Schmidt-Kaler}},\ }\bibfield
  {title} {\bibinfo {title} {Sideband cooling and coherent dynamics in a
  microchip multi-segmented ion trap},\ }\href@noop {} {\bibfield  {journal}
  {\bibinfo  {journal} {New Journal of Physics}\ }\textbf {\bibinfo {volume}
  {10}},\ \bibinfo {pages} {045007} (\bibinfo {year} {2008})}\BibitemShut
  {NoStop}%
\bibitem [{\citenamefont {Wright}\ \emph {et~al.}(2019)\citenamefont {Wright},
  \citenamefont {Beck}, \citenamefont {Debnath}, \citenamefont {Amini},
  \citenamefont {Nam}, \citenamefont {Grzesiak}, \citenamefont {Chen},
  \citenamefont {Pisenti}, \citenamefont {Chmielewski}, \citenamefont {Collins}
  \emph {et~al.}}]{wright2019benchmarking}%
  \BibitemOpen
  \bibfield  {author} {\bibinfo {author} {\bibfnamefont {K.}~\bibnamefont
  {Wright}}, \bibinfo {author} {\bibfnamefont {K.~M.}\ \bibnamefont {Beck}},
  \bibinfo {author} {\bibfnamefont {S.}~\bibnamefont {Debnath}}, \bibinfo
  {author} {\bibfnamefont {J.}~\bibnamefont {Amini}}, \bibinfo {author}
  {\bibfnamefont {Y.}~\bibnamefont {Nam}}, \bibinfo {author} {\bibfnamefont
  {N.}~\bibnamefont {Grzesiak}}, \bibinfo {author} {\bibfnamefont {J.-S.}\
  \bibnamefont {Chen}}, \bibinfo {author} {\bibfnamefont {N.}~\bibnamefont
  {Pisenti}}, \bibinfo {author} {\bibfnamefont {M.}~\bibnamefont
  {Chmielewski}}, \bibinfo {author} {\bibfnamefont {C.}~\bibnamefont
  {Collins}}, \emph {et~al.},\ }\bibfield  {title} {\bibinfo {title}
  {Benchmarking an 11-qubit quantum computer},\ }\href@noop {} {\bibfield
  {journal} {\bibinfo  {journal} {Nature communications}\ }\textbf {\bibinfo
  {volume} {10}},\ \bibinfo {pages} {1} (\bibinfo {year} {2019})}\BibitemShut
  {NoStop}%
\bibitem [{\citenamefont {Kielpinski}\ \emph {et~al.}(2002)\citenamefont
  {Kielpinski}, \citenamefont {Monroe},\ and\ \citenamefont
  {Wineland}}]{kielpinski2002architecture}%
  \BibitemOpen
  \bibfield  {author} {\bibinfo {author} {\bibfnamefont {D.}~\bibnamefont
  {Kielpinski}}, \bibinfo {author} {\bibfnamefont {C.}~\bibnamefont {Monroe}},\
  and\ \bibinfo {author} {\bibfnamefont {D.~J.}\ \bibnamefont {Wineland}},\
  }\bibfield  {title} {\bibinfo {title} {Architecture for a large-scale
  ion-trap quantum computer},\ }\href@noop {} {\bibfield  {journal} {\bibinfo
  {journal} {Nature}\ }\textbf {\bibinfo {volume} {417}},\ \bibinfo {pages}
  {709} (\bibinfo {year} {2002})}\BibitemShut {NoStop}%
\bibitem [{\citenamefont {Pino}\ \emph {et~al.}(2021)\citenamefont {Pino},
  \citenamefont {Dreiling}, \citenamefont {Figgatt}, \citenamefont {Gaebler},
  \citenamefont {Moses}, \citenamefont {Allman}, \citenamefont {Baldwin},
  \citenamefont {Foss-Feig}, \citenamefont {Hayes}, \citenamefont {Mayer} \emph
  {et~al.}}]{pino2021demonstration}%
  \BibitemOpen
  \bibfield  {author} {\bibinfo {author} {\bibfnamefont {J.~M.}\ \bibnamefont
  {Pino}}, \bibinfo {author} {\bibfnamefont {J.~M.}\ \bibnamefont {Dreiling}},
  \bibinfo {author} {\bibfnamefont {C.}~\bibnamefont {Figgatt}}, \bibinfo
  {author} {\bibfnamefont {J.~P.}\ \bibnamefont {Gaebler}}, \bibinfo {author}
  {\bibfnamefont {S.~A.}\ \bibnamefont {Moses}}, \bibinfo {author}
  {\bibfnamefont {M.}~\bibnamefont {Allman}}, \bibinfo {author} {\bibfnamefont
  {C.}~\bibnamefont {Baldwin}}, \bibinfo {author} {\bibfnamefont
  {M.}~\bibnamefont {Foss-Feig}}, \bibinfo {author} {\bibfnamefont
  {D.}~\bibnamefont {Hayes}}, \bibinfo {author} {\bibfnamefont
  {K.}~\bibnamefont {Mayer}}, \emph {et~al.},\ }\bibfield  {title} {\bibinfo
  {title} {Demonstration of the trapped-ion quantum ccd computer
  architecture},\ }\href@noop {} {\bibfield  {journal} {\bibinfo  {journal}
  {Nature}\ }\textbf {\bibinfo {volume} {592}},\ \bibinfo {pages} {209}
  (\bibinfo {year} {2021})}\BibitemShut {NoStop}%
\bibitem [{\citenamefont {Pogorelov}\ \emph {et~al.}(2021)\citenamefont
  {Pogorelov}, \citenamefont {Feldker}, \citenamefont {Marciniak},
  \citenamefont {Postler}, \citenamefont {Jacob}, \citenamefont
  {Krieglsteiner}, \citenamefont {Podlesnic}, \citenamefont {Meth},
  \citenamefont {Negnevitsky}, \citenamefont {Stadler} \emph
  {et~al.}}]{pogorelov2021compact}%
  \BibitemOpen
  \bibfield  {author} {\bibinfo {author} {\bibfnamefont {I.}~\bibnamefont
  {Pogorelov}}, \bibinfo {author} {\bibfnamefont {T.}~\bibnamefont {Feldker}},
  \bibinfo {author} {\bibfnamefont {C.~D.}\ \bibnamefont {Marciniak}}, \bibinfo
  {author} {\bibfnamefont {L.}~\bibnamefont {Postler}}, \bibinfo {author}
  {\bibfnamefont {G.}~\bibnamefont {Jacob}}, \bibinfo {author} {\bibfnamefont
  {O.}~\bibnamefont {Krieglsteiner}}, \bibinfo {author} {\bibfnamefont
  {V.}~\bibnamefont {Podlesnic}}, \bibinfo {author} {\bibfnamefont
  {M.}~\bibnamefont {Meth}}, \bibinfo {author} {\bibfnamefont {V.}~\bibnamefont
  {Negnevitsky}}, \bibinfo {author} {\bibfnamefont {M.}~\bibnamefont
  {Stadler}}, \emph {et~al.},\ }\bibfield  {title} {\bibinfo {title} {Compact
  ion-trap quantum computing demonstrator},\ }\href@noop {} {\bibfield
  {journal} {\bibinfo  {journal} {PRX Quantum}\ }\textbf {\bibinfo {volume}
  {2}},\ \bibinfo {pages} {020343} (\bibinfo {year} {2021})}\BibitemShut
  {NoStop}%
\bibitem [{\citenamefont {Rowe}\ \emph {et~al.}(2002)\citenamefont {Rowe},
  \citenamefont {Ben-Kish}, \citenamefont {Demarco}, \citenamefont {Leibfried},
  \citenamefont {Meyer}, \citenamefont {Beall}, \citenamefont {Britton},
  \citenamefont {Hughes}, \citenamefont {Itano}, \citenamefont {Jelenkovi{\'c}}
  \emph {et~al.}}]{rowe2002transport}%
  \BibitemOpen
  \bibfield  {author} {\bibinfo {author} {\bibfnamefont {M.}~\bibnamefont
  {Rowe}}, \bibinfo {author} {\bibfnamefont {A.}~\bibnamefont {Ben-Kish}},
  \bibinfo {author} {\bibfnamefont {B.}~\bibnamefont {Demarco}}, \bibinfo
  {author} {\bibfnamefont {D.}~\bibnamefont {Leibfried}}, \bibinfo {author}
  {\bibfnamefont {V.}~\bibnamefont {Meyer}}, \bibinfo {author} {\bibfnamefont
  {J.}~\bibnamefont {Beall}}, \bibinfo {author} {\bibfnamefont
  {J.}~\bibnamefont {Britton}}, \bibinfo {author} {\bibfnamefont
  {J.}~\bibnamefont {Hughes}}, \bibinfo {author} {\bibfnamefont
  {W.}~\bibnamefont {Itano}}, \bibinfo {author} {\bibfnamefont
  {B.}~\bibnamefont {Jelenkovi{\'c}}}, \emph {et~al.},\ }\bibfield  {title}
  {\bibinfo {title} {Transport of quantum states and separation of ions in a
  dual rf ion trap},\ }\href@noop {} {\bibfield  {journal} {\bibinfo  {journal}
  {Quantum Information and Computation}\ }\textbf {\bibinfo {volume} {2}},\
  \bibinfo {pages} {257} (\bibinfo {year} {2002})}\BibitemShut {NoStop}%
\bibitem [{\citenamefont {Splatt}\ \emph {et~al.}(2009)\citenamefont {Splatt},
  \citenamefont {Harlander}, \citenamefont {Brownnutt}, \citenamefont
  {Zähringer}, \citenamefont {Blatt},\ and\ \citenamefont
  {Hänsel}}]{splatt2009reordering}%
  \BibitemOpen
  \bibfield  {author} {\bibinfo {author} {\bibfnamefont {F.}~\bibnamefont
  {Splatt}}, \bibinfo {author} {\bibfnamefont {M.}~\bibnamefont {Harlander}},
  \bibinfo {author} {\bibfnamefont {M.}~\bibnamefont {Brownnutt}}, \bibinfo
  {author} {\bibfnamefont {F.}~\bibnamefont {Zähringer}}, \bibinfo {author}
  {\bibfnamefont {R.}~\bibnamefont {Blatt}},\ and\ \bibinfo {author}
  {\bibfnamefont {W.}~\bibnamefont {Hänsel}},\ }\bibfield  {title} {\bibinfo
  {title} {Deterministic reordering of $^{40}${Ca}$^+$ions in a linear
  segmented paul trap},\ }\href@noop {} {\bibfield  {journal} {\bibinfo
  {journal} {New Journal of Physics}\ }\textbf {\bibinfo {volume} {11}},\
  \bibinfo {pages} {103008} (\bibinfo {year} {2009})}\BibitemShut {NoStop}%
\bibitem [{\citenamefont {Home}\ \emph {et~al.}(2009)\citenamefont {Home},
  \citenamefont {Hanneke}, \citenamefont {Jost}, \citenamefont {Amini},
  \citenamefont {Leibfried},\ and\ \citenamefont
  {Wineland}}]{home2009complete}%
  \BibitemOpen
  \bibfield  {author} {\bibinfo {author} {\bibfnamefont {J.~P.}\ \bibnamefont
  {Home}}, \bibinfo {author} {\bibfnamefont {D.}~\bibnamefont {Hanneke}},
  \bibinfo {author} {\bibfnamefont {J.~D.}\ \bibnamefont {Jost}}, \bibinfo
  {author} {\bibfnamefont {J.~M.}\ \bibnamefont {Amini}}, \bibinfo {author}
  {\bibfnamefont {D.}~\bibnamefont {Leibfried}},\ and\ \bibinfo {author}
  {\bibfnamefont {D.~J.}\ \bibnamefont {Wineland}},\ }\bibfield  {title}
  {\bibinfo {title} {Complete methods set for scalable ion trap quantum
  information processing},\ }\href {https://doi.org/10.1126/science.1177077}
  {\bibfield  {journal} {\bibinfo  {journal} {Science}\ }\textbf {\bibinfo
  {volume} {325}},\ \bibinfo {pages} {1227} (\bibinfo {year}
  {2009})}\BibitemShut {NoStop}%
\bibitem [{\citenamefont {Hensinger}\ \emph {et~al.}(2006)\citenamefont
  {Hensinger}, \citenamefont {Olmschenk}, \citenamefont {Stick}, \citenamefont
  {Hucul}, \citenamefont {Yeo}, \citenamefont {Acton}, \citenamefont
  {Deslauriers}, \citenamefont {Monroe},\ and\ \citenamefont
  {Rabchuk}}]{hensinger2006t}%
  \BibitemOpen
  \bibfield  {author} {\bibinfo {author} {\bibfnamefont {W.}~\bibnamefont
  {Hensinger}}, \bibinfo {author} {\bibfnamefont {S.}~\bibnamefont
  {Olmschenk}}, \bibinfo {author} {\bibfnamefont {D.}~\bibnamefont {Stick}},
  \bibinfo {author} {\bibfnamefont {D.}~\bibnamefont {Hucul}}, \bibinfo
  {author} {\bibfnamefont {M.}~\bibnamefont {Yeo}}, \bibinfo {author}
  {\bibfnamefont {M.}~\bibnamefont {Acton}}, \bibinfo {author} {\bibfnamefont
  {L.}~\bibnamefont {Deslauriers}}, \bibinfo {author} {\bibfnamefont
  {C.}~\bibnamefont {Monroe}},\ and\ \bibinfo {author} {\bibfnamefont
  {J.}~\bibnamefont {Rabchuk}},\ }\bibfield  {title} {\bibinfo {title}
  {T-junction ion trap array for two-dimensional ion shuttling, storage, and
  manipulation},\ }\href@noop {} {\bibfield  {journal} {\bibinfo  {journal}
  {Applied Physics Letters}\ }\textbf {\bibinfo {volume} {88}},\ \bibinfo
  {pages} {034101} (\bibinfo {year} {2006})}\BibitemShut {NoStop}%
\bibitem [{\citenamefont {Blakestad}\ \emph {et~al.}(2009)\citenamefont
  {Blakestad}, \citenamefont {Ospelkaus}, \citenamefont {VanDevender},
  \citenamefont {Amini}, \citenamefont {Britton}, \citenamefont {Leibfried},\
  and\ \citenamefont {Wineland}}]{blakestad2009}%
  \BibitemOpen
  \bibfield  {author} {\bibinfo {author} {\bibfnamefont {R.~B.}\ \bibnamefont
  {Blakestad}}, \bibinfo {author} {\bibfnamefont {C.}~\bibnamefont
  {Ospelkaus}}, \bibinfo {author} {\bibfnamefont {A.~P.}\ \bibnamefont
  {VanDevender}}, \bibinfo {author} {\bibfnamefont {J.~M.}\ \bibnamefont
  {Amini}}, \bibinfo {author} {\bibfnamefont {J.}~\bibnamefont {Britton}},
  \bibinfo {author} {\bibfnamefont {D.}~\bibnamefont {Leibfried}},\ and\
  \bibinfo {author} {\bibfnamefont {D.~J.}\ \bibnamefont {Wineland}},\
  }\bibfield  {title} {\bibinfo {title} {High-fidelity transport of trapped-ion
  qubits through an $\mathbf{X}$-junction trap array},\ }\href
  {https://doi.org/10.1103/PhysRevLett.102.153002} {\bibfield  {journal}
  {\bibinfo  {journal} {Phys. Rev. Lett.}\ }\textbf {\bibinfo {volume} {102}},\
  \bibinfo {pages} {153002} (\bibinfo {year} {2009})}\BibitemShut {NoStop}%
\bibitem [{\citenamefont {Decaroli}\ \emph {et~al.}(2021)\citenamefont
  {Decaroli}, \citenamefont {Matt}, \citenamefont {Oswald}, \citenamefont
  {Axline}, \citenamefont {Ernzer}, \citenamefont {Flannery}, \citenamefont
  {Ragg},\ and\ \citenamefont {Home}}]{decaroli2021Zurich}%
  \BibitemOpen
  \bibfield  {author} {\bibinfo {author} {\bibfnamefont {C.}~\bibnamefont
  {Decaroli}}, \bibinfo {author} {\bibfnamefont {R.}~\bibnamefont {Matt}},
  \bibinfo {author} {\bibfnamefont {R.}~\bibnamefont {Oswald}}, \bibinfo
  {author} {\bibfnamefont {C.~J.}\ \bibnamefont {Axline}}, \bibinfo {author}
  {\bibfnamefont {M.}~\bibnamefont {Ernzer}}, \bibinfo {author} {\bibfnamefont
  {J.}~\bibnamefont {Flannery}}, \bibinfo {author} {\bibfnamefont
  {S.}~\bibnamefont {Ragg}},\ and\ \bibinfo {author} {\bibfnamefont {J.~P.}\
  \bibnamefont {Home}},\ }\bibfield  {title} {\bibinfo {title} {Design,
  fabrication and characterisation of a micro-fabricated stacked-wafer
  segmented ion trap with two x-junctions.},\ }\href@noop {} {\bibfield
  {journal} {\bibinfo  {journal} {Quantum Science and Technology}\ } (\bibinfo
  {year} {2021})}\BibitemShut {NoStop}%
\bibitem [{\citenamefont {Blakestad}\ \emph {et~al.}(2011)\citenamefont
  {Blakestad}, \citenamefont {Ospelkaus}, \citenamefont {VanDevender},
  \citenamefont {Wesenberg}, \citenamefont {Biercuk}, \citenamefont
  {Leibfried},\ and\ \citenamefont {Wineland}}]{blakestad2011xjunction}%
  \BibitemOpen
  \bibfield  {author} {\bibinfo {author} {\bibfnamefont {R.~B.}\ \bibnamefont
  {Blakestad}}, \bibinfo {author} {\bibfnamefont {C.}~\bibnamefont
  {Ospelkaus}}, \bibinfo {author} {\bibfnamefont {A.~P.}\ \bibnamefont
  {VanDevender}}, \bibinfo {author} {\bibfnamefont {J.~H.}\ \bibnamefont
  {Wesenberg}}, \bibinfo {author} {\bibfnamefont {M.~J.}\ \bibnamefont
  {Biercuk}}, \bibinfo {author} {\bibfnamefont {D.}~\bibnamefont {Leibfried}},\
  and\ \bibinfo {author} {\bibfnamefont {D.~J.}\ \bibnamefont {Wineland}},\
  }\bibfield  {title} {\bibinfo {title} {Near-ground-state transport of
  trapped-ion qubits through a multidimensional array},\ }\href
  {https://doi.org/10.1103/PhysRevA.84.032314} {\bibfield  {journal} {\bibinfo
  {journal} {Phys. Rev. A}\ }\textbf {\bibinfo {volume} {84}},\ \bibinfo
  {pages} {032314} (\bibinfo {year} {2011})}\BibitemShut {NoStop}%
\bibitem [{\citenamefont {Wright}\ \emph {et~al.}(2013)\citenamefont {Wright},
  \citenamefont {Amini}, \citenamefont {Faircloth}, \citenamefont {Volin},
  \citenamefont {Doret}, \citenamefont {Hayden}, \citenamefont {Pai},
  \citenamefont {Landgren}, \citenamefont {Denison}, \citenamefont {Killian}
  \emph {et~al.}}]{wright2013GTRIXjunction}%
  \BibitemOpen
  \bibfield  {author} {\bibinfo {author} {\bibfnamefont {K.}~\bibnamefont
  {Wright}}, \bibinfo {author} {\bibfnamefont {J.~M.}\ \bibnamefont {Amini}},
  \bibinfo {author} {\bibfnamefont {D.~L.}\ \bibnamefont {Faircloth}}, \bibinfo
  {author} {\bibfnamefont {C.}~\bibnamefont {Volin}}, \bibinfo {author}
  {\bibfnamefont {S.~C.}\ \bibnamefont {Doret}}, \bibinfo {author}
  {\bibfnamefont {H.}~\bibnamefont {Hayden}}, \bibinfo {author} {\bibfnamefont
  {C.}~\bibnamefont {Pai}}, \bibinfo {author} {\bibfnamefont {D.~W.}\
  \bibnamefont {Landgren}}, \bibinfo {author} {\bibfnamefont {D.}~\bibnamefont
  {Denison}}, \bibinfo {author} {\bibfnamefont {T.}~\bibnamefont {Killian}},
  \emph {et~al.},\ }\bibfield  {title} {\bibinfo {title} {Reliable transport
  through a microfabricated x-junction surface-electrode ion trap},\
  }\href@noop {} {\bibfield  {journal} {\bibinfo  {journal} {New Journal of
  Physics}\ }\textbf {\bibinfo {volume} {15}},\ \bibinfo {pages} {033004}
  (\bibinfo {year} {2013})}\BibitemShut {NoStop}%
\bibitem [{\citenamefont {Zhang}\ \emph {et~al.}(2022)\citenamefont {Zhang},
  \citenamefont {Mehta},\ and\ \citenamefont {Home}}]{zhang2022optimization}%
  \BibitemOpen
  \bibfield  {author} {\bibinfo {author} {\bibfnamefont {C.}~\bibnamefont
  {Zhang}}, \bibinfo {author} {\bibfnamefont {K.~K.}\ \bibnamefont {Mehta}},\
  and\ \bibinfo {author} {\bibfnamefont {J.~P.}\ \bibnamefont {Home}},\ }\href
  {https://doi.org/10.48550/ARXIV.2201.12579} {\bibinfo {title} {Optimization
  and implementation of a surface-electrode ion trap junction}} (\bibinfo
  {year} {2022}),\ \Eprint {https://arxiv.org/abs/2201.12579}
  {arXiv:2201.12579} \BibitemShut {NoStop}%
\bibitem [{\citenamefont {Amini}\ \emph {et~al.}(2010)\citenamefont {Amini},
  \citenamefont {Uys}, \citenamefont {Wesenberg}, \citenamefont {Seidelin},
  \citenamefont {Britton}, \citenamefont {Bollinger}, \citenamefont
  {Leibfried}, \citenamefont {Ospelkaus}, \citenamefont {VanDevender},\ and\
  \citenamefont {Wineland}}]{amini2010scalabletraps}%
  \BibitemOpen
  \bibfield  {author} {\bibinfo {author} {\bibfnamefont {J.~M.}\ \bibnamefont
  {Amini}}, \bibinfo {author} {\bibfnamefont {H.}~\bibnamefont {Uys}}, \bibinfo
  {author} {\bibfnamefont {J.~H.}\ \bibnamefont {Wesenberg}}, \bibinfo {author}
  {\bibfnamefont {S.}~\bibnamefont {Seidelin}}, \bibinfo {author}
  {\bibfnamefont {J.}~\bibnamefont {Britton}}, \bibinfo {author} {\bibfnamefont
  {J.~J.}\ \bibnamefont {Bollinger}}, \bibinfo {author} {\bibfnamefont
  {D.}~\bibnamefont {Leibfried}}, \bibinfo {author} {\bibfnamefont
  {C.}~\bibnamefont {Ospelkaus}}, \bibinfo {author} {\bibfnamefont {A.~P.}\
  \bibnamefont {VanDevender}},\ and\ \bibinfo {author} {\bibfnamefont {D.~J.}\
  \bibnamefont {Wineland}},\ }\bibfield  {title} {\bibinfo {title} {Toward
  scalable ion traps for quantum information processing},\ }\href@noop {}
  {\bibfield  {journal} {\bibinfo  {journal} {New journal of Physics}\ }\textbf
  {\bibinfo {volume} {12}},\ \bibinfo {pages} {033031} (\bibinfo {year}
  {2010})}\BibitemShut {NoStop}%
\bibitem [{\citenamefont {Moehring}\ \emph {et~al.}(2011)\citenamefont
  {Moehring}, \citenamefont {Highstrete}, \citenamefont {Stick}, \citenamefont
  {Fortier}, \citenamefont {Haltli}, \citenamefont {Tigges},\ and\
  \citenamefont {Blain}}]{moehring2011SandiaJunction}%
  \BibitemOpen
  \bibfield  {author} {\bibinfo {author} {\bibfnamefont {D.~L.}\ \bibnamefont
  {Moehring}}, \bibinfo {author} {\bibfnamefont {C.}~\bibnamefont
  {Highstrete}}, \bibinfo {author} {\bibfnamefont {D.}~\bibnamefont {Stick}},
  \bibinfo {author} {\bibfnamefont {K.~M.}\ \bibnamefont {Fortier}}, \bibinfo
  {author} {\bibfnamefont {R.}~\bibnamefont {Haltli}}, \bibinfo {author}
  {\bibfnamefont {C.}~\bibnamefont {Tigges}},\ and\ \bibinfo {author}
  {\bibfnamefont {M.~G.}\ \bibnamefont {Blain}},\ }\bibfield  {title} {\bibinfo
  {title} {Design, fabrication and experimental demonstration of junction
  surface ion traps},\ }\href@noop {} {\bibfield  {journal} {\bibinfo
  {journal} {New Journal of Physics}\ }\textbf {\bibinfo {volume} {13}},\
  \bibinfo {pages} {075018} (\bibinfo {year} {2011})}\BibitemShut {NoStop}%
\bibitem [{\citenamefont {Shu}\ \emph {et~al.}(2014)\citenamefont {Shu},
  \citenamefont {Vittorini}, \citenamefont {Buikema}, \citenamefont {Nichols},
  \citenamefont {Volin}, \citenamefont {Stick},\ and\ \citenamefont
  {Brown}}]{shu2014GTRIYjunction}%
  \BibitemOpen
  \bibfield  {author} {\bibinfo {author} {\bibfnamefont {G.}~\bibnamefont
  {Shu}}, \bibinfo {author} {\bibfnamefont {G.}~\bibnamefont {Vittorini}},
  \bibinfo {author} {\bibfnamefont {A.}~\bibnamefont {Buikema}}, \bibinfo
  {author} {\bibfnamefont {C.}~\bibnamefont {Nichols}}, \bibinfo {author}
  {\bibfnamefont {C.}~\bibnamefont {Volin}}, \bibinfo {author} {\bibfnamefont
  {D.}~\bibnamefont {Stick}},\ and\ \bibinfo {author} {\bibfnamefont {K.~R.}\
  \bibnamefont {Brown}},\ }\bibfield  {title} {\bibinfo {title} {Heating rates
  and ion-motion control in a y-junction surface-electrode trap},\ }\href@noop
  {} {\bibfield  {journal} {\bibinfo  {journal} {Physical Review A}\ }\textbf
  {\bibinfo {volume} {89}},\ \bibinfo {pages} {062308} (\bibinfo {year}
  {2014})}\BibitemShut {NoStop}%
\bibitem [{\citenamefont {Drees}\ and\ \citenamefont
  {Paul}(1964)}]{drees1964beschleunigung}%
  \BibitemOpen
  \bibfield  {author} {\bibinfo {author} {\bibfnamefont {J.}~\bibnamefont
  {Drees}}\ and\ \bibinfo {author} {\bibfnamefont {W.}~\bibnamefont {Paul}},\
  }\bibfield  {title} {\bibinfo {title} {Beschleunigung von elektronen in einem
  plasmabetatron},\ }\href@noop {} {\bibfield  {journal} {\bibinfo  {journal}
  {Zeitschrift f{\"u}r Physik}\ }\textbf {\bibinfo {volume} {180}},\ \bibinfo
  {pages} {340} (\bibinfo {year} {1964})}\BibitemShut {NoStop}%
\bibitem [{\citenamefont {Hucul}\ \emph {et~al.}(2008)\citenamefont {Hucul},
  \citenamefont {Yeo}, \citenamefont {Olmschenk}, \citenamefont {Monroe},
  \citenamefont {Hensinger},\ and\ \citenamefont
  {Rabchuk}}]{hucul2008transport}%
  \BibitemOpen
  \bibfield  {author} {\bibinfo {author} {\bibfnamefont {D.}~\bibnamefont
  {Hucul}}, \bibinfo {author} {\bibfnamefont {M.}~\bibnamefont {Yeo}}, \bibinfo
  {author} {\bibfnamefont {S.}~\bibnamefont {Olmschenk}}, \bibinfo {author}
  {\bibfnamefont {C.}~\bibnamefont {Monroe}}, \bibinfo {author} {\bibfnamefont
  {W.}~\bibnamefont {Hensinger}},\ and\ \bibinfo {author} {\bibfnamefont
  {J.}~\bibnamefont {Rabchuk}},\ }\bibfield  {title} {\bibinfo {title} {On the
  transport of atomic ions in linear and multidimensional ion trap arrays},\
  }\href@noop {} {\bibfield  {journal} {\bibinfo  {journal} {Quantum
  Information \& Computation}\ }\textbf {\bibinfo {volume} {8}},\ \bibinfo
  {pages} {501} (\bibinfo {year} {2008})}\BibitemShut {NoStop}%
\bibitem [{\citenamefont {Wesenberg}(2009)}]{wesenberg2009idealintersections}%
  \BibitemOpen
  \bibfield  {author} {\bibinfo {author} {\bibfnamefont {J.~H.}\ \bibnamefont
  {Wesenberg}},\ }\bibfield  {title} {\bibinfo {title} {Ideal intersections for
  radio-frequency trap networks},\ }\href
  {https://doi.org/10.1103/PhysRevA.79.013416} {\bibfield  {journal} {\bibinfo
  {journal} {Phys. Rev. A}\ }\textbf {\bibinfo {volume} {79}},\ \bibinfo
  {pages} {013416} (\bibinfo {year} {2009})}\BibitemShut {NoStop}%
\bibitem [{\citenamefont {Monroe}\ \emph {et~al.}(1995)\citenamefont {Monroe},
  \citenamefont {Meekhof}, \citenamefont {King}, \citenamefont {Jefferts},
  \citenamefont {Itano}, \citenamefont {Wineland},\ and\ \citenamefont
  {Gould}}]{PhysRevLett.75.4011}%
  \BibitemOpen
  \bibfield  {author} {\bibinfo {author} {\bibfnamefont {C.}~\bibnamefont
  {Monroe}}, \bibinfo {author} {\bibfnamefont {D.~M.}\ \bibnamefont {Meekhof}},
  \bibinfo {author} {\bibfnamefont {B.~E.}\ \bibnamefont {King}}, \bibinfo
  {author} {\bibfnamefont {S.~R.}\ \bibnamefont {Jefferts}}, \bibinfo {author}
  {\bibfnamefont {W.~M.}\ \bibnamefont {Itano}}, \bibinfo {author}
  {\bibfnamefont {D.~J.}\ \bibnamefont {Wineland}},\ and\ \bibinfo {author}
  {\bibfnamefont {P.}~\bibnamefont {Gould}},\ }\bibfield  {title} {\bibinfo
  {title} {Resolved-sideband raman cooling of a bound atom to the 3d zero-point
  energy},\ }\href {https://doi.org/10.1103/PhysRevLett.75.4011} {\bibfield
  {journal} {\bibinfo  {journal} {Phys. Rev. Lett.}\ }\textbf {\bibinfo
  {volume} {75}},\ \bibinfo {pages} {4011} (\bibinfo {year}
  {1995})}\BibitemShut {NoStop}%
\bibitem [{\citenamefont {Palmero}\ \emph {et~al.}(2014)\citenamefont
  {Palmero}, \citenamefont {Bowler}, \citenamefont {Gaebler}, \citenamefont
  {Leibfried},\ and\ \citenamefont {Muga}}]{palmero2014fasttransport}%
  \BibitemOpen
  \bibfield  {author} {\bibinfo {author} {\bibfnamefont {M.}~\bibnamefont
  {Palmero}}, \bibinfo {author} {\bibfnamefont {R.}~\bibnamefont {Bowler}},
  \bibinfo {author} {\bibfnamefont {J.~P.}\ \bibnamefont {Gaebler}}, \bibinfo
  {author} {\bibfnamefont {D.}~\bibnamefont {Leibfried}},\ and\ \bibinfo
  {author} {\bibfnamefont {J.~G.}\ \bibnamefont {Muga}},\ }\bibfield  {title}
  {\bibinfo {title} {Fast transport of mixed-species ion chains within a paul
  trap},\ }\href {https://doi.org/10.1103/PhysRevA.90.053408} {\bibfield
  {journal} {\bibinfo  {journal} {Phys. Rev. A}\ }\textbf {\bibinfo {volume}
  {90}},\ \bibinfo {pages} {053408} (\bibinfo {year} {2014})}\BibitemShut
  {NoStop}%
\end{thebibliography}%


\begin{thebibliography}{2}%
\makeatletter
\providecommand \@ifxundefined [1]{%
 \@ifx{#1\undefined}
}%
\providecommand \@ifnum [1]{%
 \ifnum #1\expandafter \@firstoftwo
 \else \expandafter \@secondoftwo
 \fi
}%
\providecommand \@ifx [1]{%
 \ifx #1\expandafter \@firstoftwo
 \else \expandafter \@secondoftwo
 \fi
}%
\providecommand \natexlab [1]{#1}%
\providecommand \enquote  [1]{``#1''}%
\providecommand \bibnamefont  [1]{#1}%
\providecommand \bibfnamefont [1]{#1}%
\providecommand \citenamefont [1]{#1}%
\providecommand \href@noop [0]{\@secondoftwo}%
\providecommand \href [0]{\begingroup \@sanitize@url \@href}%
\providecommand \@href[1]{\@@startlink{#1}\@@href}%
\providecommand \@@href[1]{\endgroup#1\@@endlink}%
\providecommand \@sanitize@url [0]{\catcode `\\12\catcode `\$12\catcode
  `\&12\catcode `\#12\catcode `\^12\catcode `\_12\catcode `\%12\relax}%
\providecommand \@@startlink[1]{}%
\providecommand \@@endlink[0]{}%
\providecommand \url  [0]{\begingroup\@sanitize@url \@url }%
\providecommand \@url [1]{\endgroup\@href {#1}{\urlprefix }}%
\providecommand \urlprefix  [0]{URL }%
\providecommand \Eprint [0]{\href }%
\providecommand \doibase [0]{https://doi.org/}%
\providecommand \selectlanguage [0]{\@gobble}%
\providecommand \bibinfo  [0]{\@secondoftwo}%
\providecommand \bibfield  [0]{\@secondoftwo}%
\providecommand \translation [1]{[#1]}%
\providecommand \BibitemOpen [0]{}%
\providecommand \bibitemStop [0]{}%
\providecommand \bibitemNoStop [0]{.\EOS\space}%
\providecommand \EOS [0]{\spacefactor3000\relax}%
\providecommand \BibitemShut  [1]{\csname bibitem#1\endcsname}%
\let\auto@bib@innerbib\@empty
\bibitem [{\citenamefont {Pino}\ \emph {et~al.}(2021)\citenamefont {Pino},
  \citenamefont {Dreiling}, \citenamefont {Figgatt}, \citenamefont {Gaebler},
  \citenamefont {Moses}, \citenamefont {Allman}, \citenamefont {Baldwin},
  \citenamefont {Foss-Feig}, \citenamefont {Hayes}, \citenamefont {Mayer} \emph
  {et~al.}}]{pino2021demonstration}%
  \BibitemOpen
  \bibfield  {author} {\bibinfo {author} {\bibfnamefont {J.~M.}\ \bibnamefont
  {Pino}}, \bibinfo {author} {\bibfnamefont {J.~M.}\ \bibnamefont {Dreiling}},
  \bibinfo {author} {\bibfnamefont {C.}~\bibnamefont {Figgatt}}, \bibinfo
  {author} {\bibfnamefont {J.~P.}\ \bibnamefont {Gaebler}}, \bibinfo {author}
  {\bibfnamefont {S.~A.}\ \bibnamefont {Moses}}, \bibinfo {author}
  {\bibfnamefont {M.}~\bibnamefont {Allman}}, \bibinfo {author} {\bibfnamefont
  {C.}~\bibnamefont {Baldwin}}, \bibinfo {author} {\bibfnamefont
  {M.}~\bibnamefont {Foss-Feig}}, \bibinfo {author} {\bibfnamefont
  {D.}~\bibnamefont {Hayes}}, \bibinfo {author} {\bibfnamefont
  {K.}~\bibnamefont {Mayer}}, \emph {et~al.},\ }\bibfield  {title} {\bibinfo
  {title} {Demonstration of the trapped-ion quantum ccd computer
  architecture},\ }\href@noop {} {\bibfield  {journal} {\bibinfo  {journal}
  {Nature}\ }\textbf {\bibinfo {volume} {592}},\ \bibinfo {pages} {209}
  (\bibinfo {year} {2021})}\BibitemShut {NoStop}%
\bibitem [{\citenamefont {Berkeland}\ \emph {et~al.}(1998)\citenamefont
  {Berkeland}, \citenamefont {Miller}, \citenamefont {Bergquist}, \citenamefont
  {Itano},\ and\ \citenamefont {Wineland}}]{berkeland1998micromotion}%
  \BibitemOpen
  \bibfield  {author} {\bibinfo {author} {\bibfnamefont {D.~J.}\ \bibnamefont
  {Berkeland}}, \bibinfo {author} {\bibfnamefont {J.~D.}\ \bibnamefont
  {Miller}}, \bibinfo {author} {\bibfnamefont {J.~C.}\ \bibnamefont
  {Bergquist}}, \bibinfo {author} {\bibfnamefont {W.~M.}\ \bibnamefont
  {Itano}},\ and\ \bibinfo {author} {\bibfnamefont {D.~J.}\ \bibnamefont
  {Wineland}},\ }\bibfield  {title} {\bibinfo {title} {Minimization of ion
  micromotion in a paul trap},\ }\href {https://doi.org/10.1063/1.367318}
  {\bibfield  {journal} {\bibinfo  {journal} {Journal of Applied Physics}\
  }\textbf {\bibinfo {volume} {83}},\ \bibinfo {pages} {5025} (\bibinfo {year}
  {1998})},\ \Eprint {https://arxiv.org/abs/https://doi.org/10.1063/1.367318}
  {https://doi.org/10.1063/1.367318} \BibitemShut {NoStop}%
\end{thebibliography}%

\end{document}


\title{Supplemental material for\\ Transport of multispecies ion crystals through a junction in an RF Paul trap}

\author{William Cody Burton}
\email{william.burton@quantinuum.com}

\author{Brian Estey}
\author{Ian M. Hoffman}
\author{Abigail R. Perry}
\author{Curtis Volin}
\author{Gabriel Price}
\email{gabriel.price@quantinuum.com}

\affiliation{Quantinuum, 303 S. Technology Ct., Broomfield, Colorado 80021, USA}


\date{\today}
\maketitle

\section{Multispecies Junction Transport Optimization}

\begin{figure}
\includegraphics[width=\columnwidth]{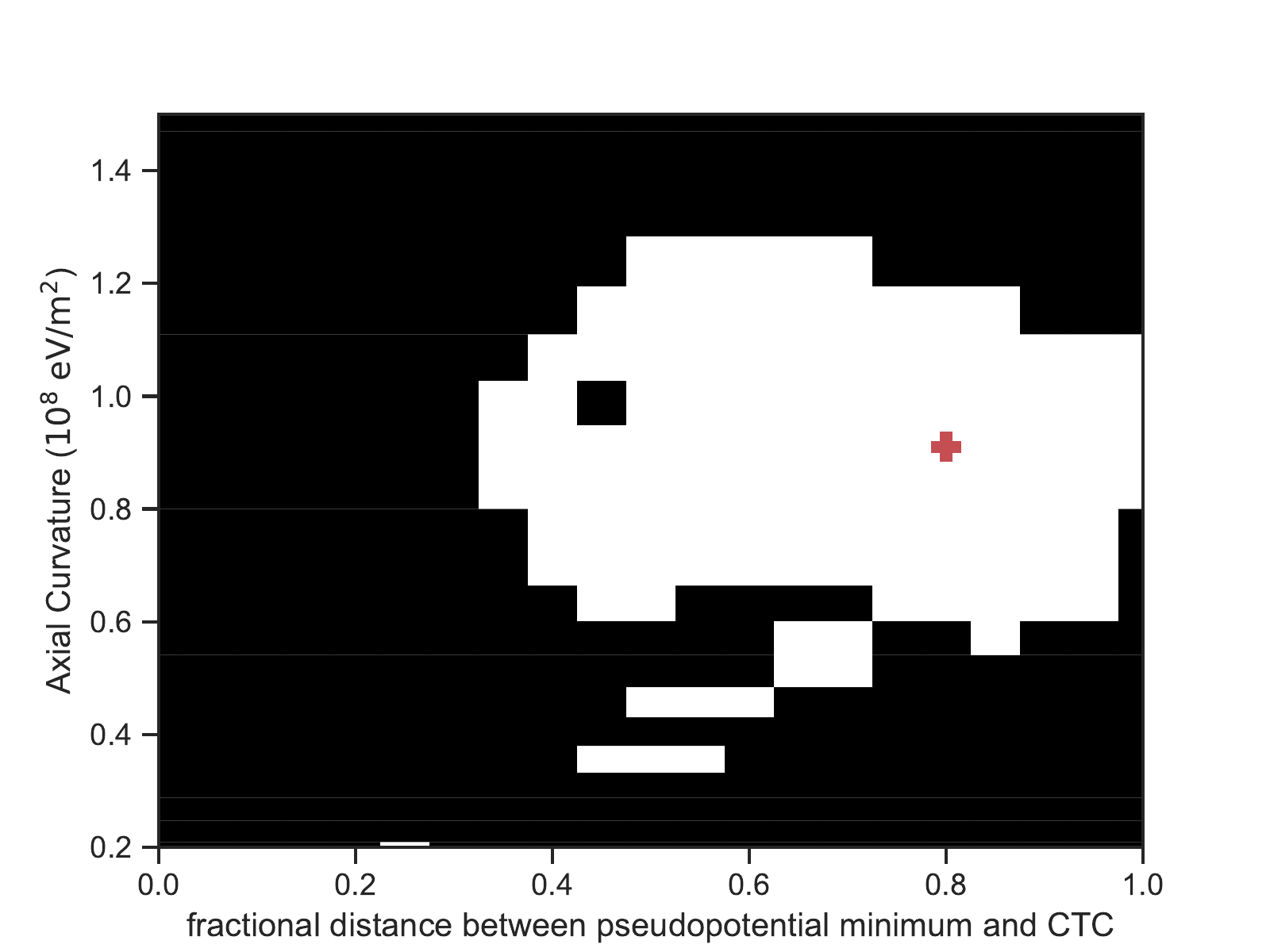}
\caption{\label{gridsearch_excitation_fig} Plot of results with excitation less than (white) or greater than (black) 10 quanta during numerical optimization. The parameters chosen for this paper are marked with a red +.}
\end{figure}

In the process of creating transport waveforms for a Ba-Yb crystal, we create test waveforms assuming a single synthetic ion with a mass equal to the average mass in the Ba-Yb crystal. 
As discussed in the main text, each test waveform is parametrized by the curvature of the total potential in the axial direction and by the ion path, which is given as a weighted combination of the paths of minimum pseudopotential and CTC. 
In Figure \ref{gridsearch_excitation_fig}, we plot the non-adiabatic excitation predicted by a numerical equations-of-motion solver for a Ba-Yb crystal after applying each test waveform, with results of greater than ten quanta of excitation colored black and less than 10 quanta colored white. The waveform used in the main paper (marked with a red + in the plot) is far from the edges of the region of low excitation.

\section{Sensitivity Analysis}

\begin{figure}
\includegraphics[width=\columnwidth]{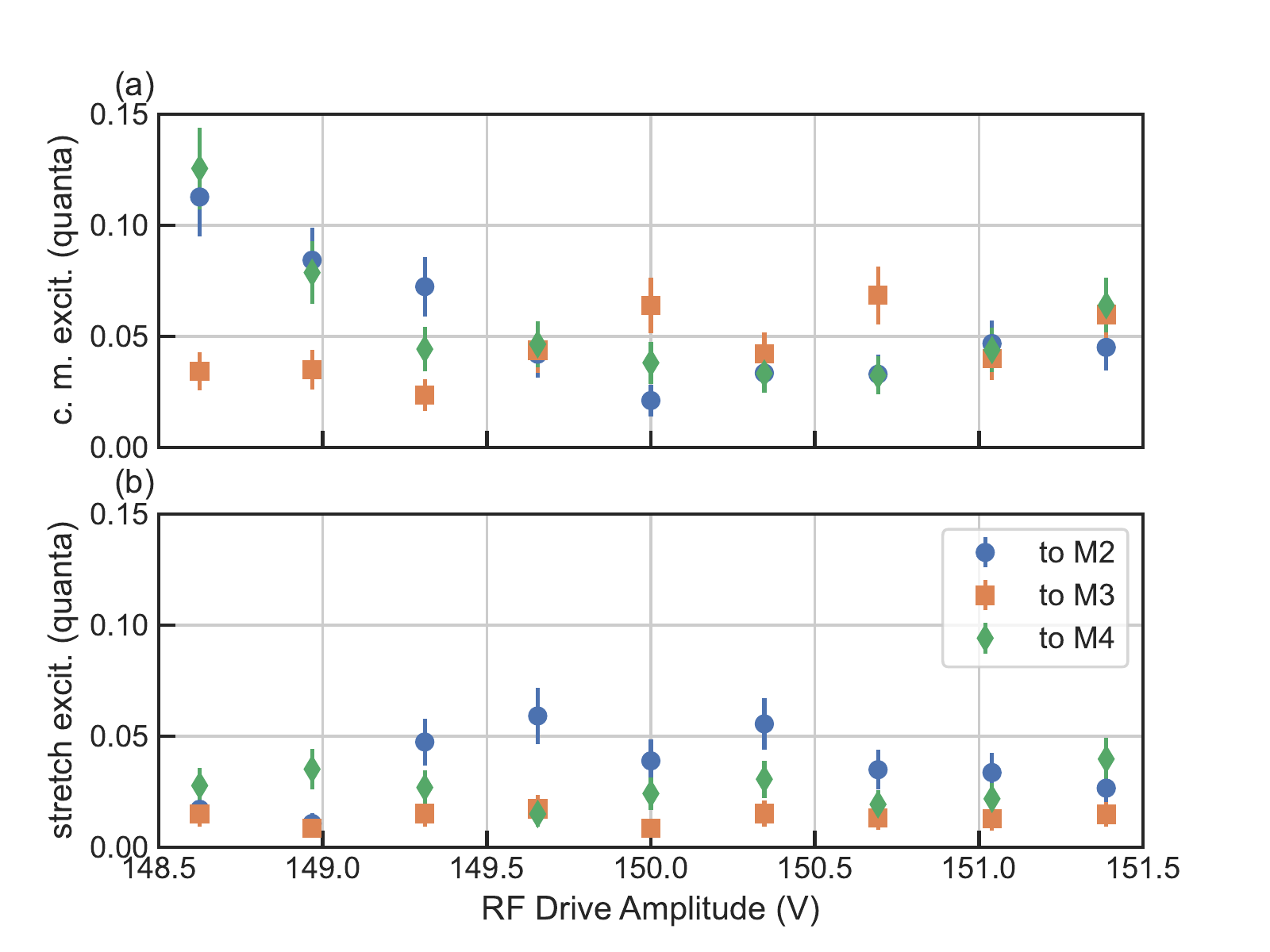}
\caption{\label{rf_fig}(a) Center of mass (c.\,m.) and (b) stretch mode sensitivity to drifts in the RF drive amplitude when Ba-Yb junction transport is performed to zone M2 (blue circles), M3 (orange squares), and M4 (green diamonds).}
\end{figure}

Understanding the sensitivity of our waveform generation method to perturbing system drifts helps us estimate the needed recalibration frequency~\cite{pino2021demonstration}. 
Therefore, we measured the sensitivity of our
transport method to two major perturbations known to impact ion survivability and transport induced excitation. First, we varied the amplitude of the RF drive for the RF source over a range of $\pm 1\%$ around its nominal value of $150\;\mathrm{V}$, while holding its frequency constant at $44.3\;\mathrm{MHz}$. The resulting axial excitation after one round-trip at an average speed of 4 m/s is plotted in Figure \ref{rf_fig}. We find that the final excitation is essentially independent of RF voltage at this level. 

\begin{figure}
\includegraphics[width=\columnwidth]{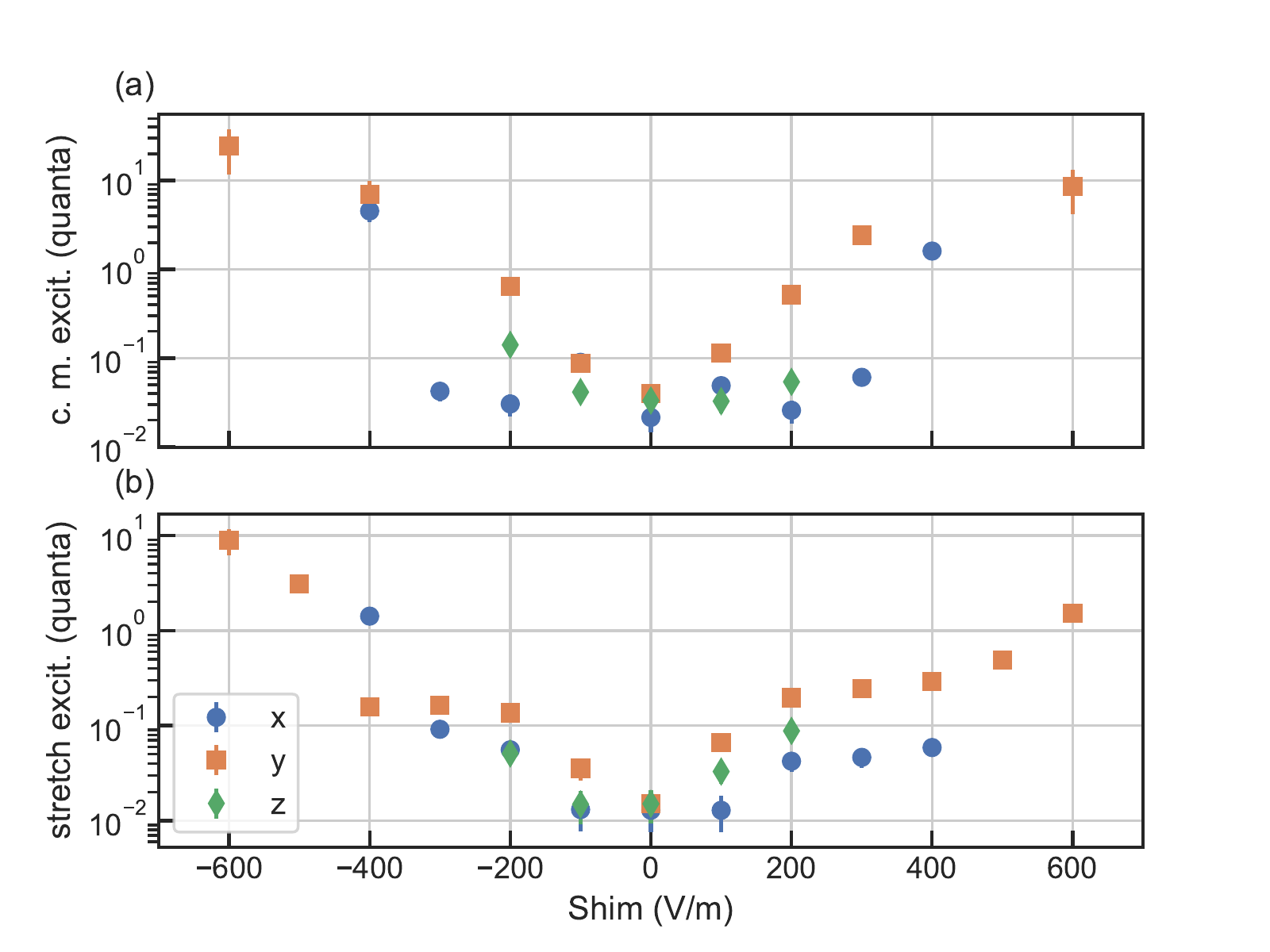}
\caption{\label{shim_fig}The sensitivity of the c.\,m. and stretch mode axial excitation to changes in the bias electric field along the $\hat{x}$ (blue circles), $\hat{y}$ (orange squares), and $\hat{z}$ (green diamonds) directions.}
\end{figure}

Additionally, we measured the impact of stray electric fields~\cite{berkeland1998micromotion} on our transport method by adding artificial ``stray fields" to the transport waveform.
At each step of the waveform, we calculated three sets of voltages 
that give a $1\;\mathrm{V/m}$ electric field in $\hat{x}$, $\hat{y}$, and $\hat{z}$, with no curvature, at the equilibrium position of the ion crystal. 
We scale and add these stray field waveforms to the transport waveform, and then measure the excitation after a round trip as described in the main paper. 
In Figure \ref{shim_fig}, we plot the excitation versus applied stray field in each principle direction. For some large applied stray fields, either the 
crystal becomes untrapped, or the transport induces large coherent oscillation, making sideband asymmetry a poor measure. In those cases,
we do not plot any points. The crystal survives with low coherent motion out to $\pm 400\;\mathrm{V/m}$, $\pm 600\;\mathrm{V/m}$, and 
$\pm 200\;\mathrm{V/m}$ for stray fields in the $\hat{x}$, $\hat{y}$, and $\hat{z}$ directions respectively. For all directions, the additional excitation 
is less than 1 quanta for fields smaller than $\pm 200\;\mathrm{V/m}$. As a point of reference, typical electrical field drift rates in our system is on the order of $\sim10\;\mathrm{V/m}$ over a day, indicating that this waveform generation method is robust against typical drifts.

\bibliography{supplement_refs}